# A Novel Method of Restoration Path Optimization for the AC-DC Bulk Power Grid after a Major Blackout

Chao Yang[1*], Gaoshen Liang[2], Tianle Cheng[3], Yang Li[4], Shaoyan Li[1]

[1] School of Electrical and Electronic Engineering, North China Electric Power University, Baoding, China
[2] School of Information Technology, Beijing Normal University, Zhuhai, Zhuhai, China
[3] School of Life and Health Science, The Chinese University of Hong Kong, Shenzhen, Shenzhen, China
[4] School of Electrical Engineering, Northeast Electric Power University, Jilin, China
[*] yangchao@cuhk.edu.cn

**Abstract**: The restoration control of the modern alternating current-direct current (AC-DC) hybrid power grid after a major blackout is difficult and complex. Taking into account the interaction between the line-commutated converter high-voltage direct current (LCC-HVDC) and the AC power grid, this paper proposes a novel optimization method of restoration path to reconfigure the skeleton network for the blackout power grid. Based on the system strength, the supporting capability of the AC power grid for the LCC-HVDC is firstly analysed from the aspects of start-up and operation of LCC-HVDCs. Subsequently, the quantitative relationship between the restoration path and the restoration characteristic of LCC-HVDC is derived in detail based on the system strength indices of the short circuit capacity and the frequency regulation capability. Then, an optimization model of restoration path considering non-tree paths is formulated and a feasible optimization algorithm is proposed to achieve the optimal path restoration scheme. A modified IEEE 39-bus system and a partial power grid of Southwest China are simulated to show that the proposed method is suitable for the restoration of AC-DC power grids and can improve restoration efficiency. This research can be an important guidance for operators to rapidly restore the AC-DC power grid.

**Nomenclature**

*Abbreviations*

| | |
|---|---|
| AC | Alternating current |
| DC | Direct current |
| HVDC | High-voltage direct current |
| LCC-HVDC | Line-commutated converter HVDC |
| FRC | Frequency regulation capability |
| SCC | Short circuit capacity |
| SCR | Short circuit ratio |
| ITS-MPGA | Improved topological sorting and multi-population genetic algorithm |
| PRS | Path restoration scheme |

*Indices*

| | |
|---|---|
| $d$ | Index of nodes with LCC-HVDCs |
| $g, N_G$ | Index and total number of generators |
| $i, j, N_B$ | Index and total number of nodes |
| $l, N_L$ | Index and total number of branches |

*Parameters*

| | |
|---|---|
| $E_{LL}$ | Effective value of the transformer in the inverter station |
| $df_g$ | Transient frequency response coefficient of generator $g$ |
| $\Delta f_{max}$ | Maximum frequency deviation of power grids |
| $M_f^{min}$ | Lower limit of FRC |
| $N_r$ | Number of bridges for LCC-HVDCs |
| $P_{cr,g}$ | Start-up power requirement of generator $g$ |
| $P_{DN}$ | Rated DC transmission power |
| $P_D^{min}$ | Minimum start-up power of LCC-HVDCs |
| $P_{DL}$ | Power consumption of the inverter station |
| $P_{GN,g}$ | Rated active power of generator $g$ |
| $P_{G,g}^{min}, P_{G,g}^{max}$ | Lower and upper active power limits of generator $g$ |
| $P_{L,l}^{min}, P_{L,l}^{max}$ | Lower and upper active power limits of branch $l$ |
| $Q_{A,g}^{max}$ | Maximum reactive power generator $g$ can absorb |
| $Q_F$ | Minimum filter capacity in inverter stations |
| $Q_{G,g}^{min}, Q_{G,g}^{max}$ | Lower and upper reactive power limits of generator $g$ |
| $r_g$ | Ramping rate of generator $g$ |
| $S_{GN,g}$ | Rated capacity of generator $g$ |
| $S_{sc}^{min}$ | Lower limit of short circuit capacity |
| $T_{CC,g}$ | Minimum critical time of cold start-up |
| $T_{CH,g}$ | Maximum critical time of hot start-up |
| $U_{B,i}^{min}, U_{B,i}^{max}$ | Lower and upper voltage limits of node $i$ |
| $U_{D0}$ | Ideal no-load DC voltage of inverter |
| $U_N$ | Rated voltage of nodes |
| $\Delta U_{max}$ | Maximum voltage deviation |
| $x'_g$ | Transient reactance of generator $g$ |
| $X_r$ | Reactance of transformer in inverter stations |
| $z_{\alpha\alpha}$ | Impedance of branch $\alpha$ |

*Variables*

| | |
|---|---|
| $I_D$ | HVDC current |
| $K_{CB,g}$ | Short circuit ratio of generator $g$ |
| $M_f$ | FRC of restored power systems |
| $P_{B,i}$ | Restored load of node $i$ |
| $P_D$ | HVDC transmission power |
| $P_D^{up}$ | Upper limit of DC transmission power |
| $P_{G,g}$ | Active power of generator $g$ |
| $P_{L,l}$ | Active power flowing through branch $l$ |
| $Q_{B,i}$ | Reactive power of node $i$ |
| $Q_D$ | Reactive power of the inverter |
| $Q_{G,g}$ | Reactive power of generator $g$ |
| $Q_{L,l}$ | Charging reactive power of branch $l$ |
| $S_{CR}$ | Short circuit ratio of nodes |
| $S_{sc}$ | Short circuit capacity |
| $t_{start}$ | Start-up time of the LCC-HVDC |
| $t_1, t_m, t_{end}$ | First, m-th and end time when $P_D$ increases |
| $t'_{s,g}$ | Start-up time of generator $g$ |
| $t'_{s,g}$ | Grid-connection time of generator $g$ |
| $t''_{s,g}$ | Time of reaching rated power of generator $g$ |
| $T$ | Total restoration time |
| $U_B$ | Voltage of nodes |
| $U_D$ | HVDC voltage |
| $z_{dd}, z'_{dd}$ | Thevenin equivalent impedance of node $d$ |
| $z_{dp}, z_{dq}$ | Mutual-impedance between nodes $d$ and $p$, $d$ and $q$ |
| $z_{pp}, z_{qq}$ | Self-impedance of nodes $p$ and $q$ |
| $z_{pq}, z_{qp}$ | Mutual-impedance between nodes $p$ and $q$ |
| $\gamma$ | Extinction angle of the inverter |
| $\delta_g, \delta_i, \delta_l$ | Bool variable representing the restoration state of generator $g$, node $i$ and branch $l$ |
| $\theta_{ij}$ | Phase difference between nodes $i$ and $j$ |

## 1. Introduction

A main feature of the modern power grid is the AC system and DC system hybrid connection [1],[2]. The high-voltage direct current (HVDC) changes the operation characteristics of power grids [3], and it is more complex to control the AC-DC hybrid power grid. When the power grid blackouts, the HVDC needs to deeply participate in the restoration, which makes new changes in the AC-DC power grid restoration. Therefore, the research on the restoration strategy of the AC-DC power grid has important practical value.

At present, the LCC-HVDC has been widely used in the transmission grid [4]. After a major blackout, operators usually shut down the HVDCs to prevent the extension of the blackout. HVDCs generally act as external power sources when restoring the AC-DC power grid [5]. Thus, it is of great significance to study how to reasonably utilize HVDCs to accelerate the restoration process.

In the restoration process of the traditional AC power grid, the main role of the restoration path [6] is to connect generators and loads and to transmit power between them. The restoration speed can be significantly improved by reasonably optimizing the path restoration sequence [7].

The current research on restoration paths is mostly for the AC grid. There are mainly two kinds of methods: the graph theory-based method [8]-[10] and the optimization model-based method [6],[11],[12]. The graph theory-based methods use classical graph algorithms to search the path between targets, such as the shortest path (SP) algorithm [13],[14] and the minimum spanning tree algorithm [15]. The optimization model-based methods use mathematical programming [11],[16] or artificial intelligence algorithms [9] to search optimal restoration path schemes. These existing researches aim at rapid restoration and believe that path optimization is the spanning tree problem of a graph. So their restoration schemes usually form a tree network [17] with some good characteristics. These characteristics include the maximum generation power [11], the optimal weighted sum of restoration paths [6] and the synthesis of multiple objectives [18]. Conversely, the non-tree path increases the redundancy of restoration paths, which usually results in longer restoration time. This is contrary to the traditional rapid restoration of AC power grids. Thus, there is little research on the non-tree path problem with loop networks. However, loop networks can improve the resilience of restored power grids in certain aspects. Ref. [19] constructs loop networks to eliminate line overloads in the restoration process. But it only focuses on the moment when overloads occur, rather than the whole process of path restoration.

Some researchers have studied the power grid restoration considering the LCC-HVDC. Ref. [20] studies the optimal start-up mode and control strategy of the LCC-HVDC by simulation. Ref. [21] analyzes the impact of the LCC-HVDC start-up on the AC grid and the requirement for system strength. They are only concerned about the start-up problems of LCC-HVDC itself, not the restoration of the AC-DC hybrid power grid. Ref. [24] clarifies the restoration relationship between the LCC-HVDC and the AC grid. It illustrates the quantitative relationship between the restoration path and system strength. However, these existing research are mainly focused on the start-up of LCC-HVDCs. There is insufficient research on how to improve LCC-HVDC support power by restoring the power grid and how LCC-HVDC participates in the whole restoration process. Ref. [5] considers the participation of multiple LCC-HVDCs in formulating partitioning restoration strategies. Refs. [22] and [23] establish optimization models of restoration paths with the HVDC participation. Ref. [25] proposes a parallel restoration method for AC-DC hybrid power systems. However, these studies still use the SP algorithm to search the restoration path, which leads to the underutilization of LCC-HVDCs' power.

The restoration path directly represents the restoration state of the power grid and then impacts the system strength. Meanwhile, the system strength determines the operation state of LCC-HVDCs. Therefore, the restoration path has a new function of supporting the restoration of LCC-HVDC. The system strength of the restored AC power grid is jointly determined by the topology and power grid parameters [24]. The restoration operations that mainly affect the system strength include the restoration of generators and loop networks. Hence, reasonable restoration paths can increase the system strength and provide strong support for LCC-HVDCs.

Generally, the power regulation of LCC-HVDCs is very fast. It only takes hundreds of milliseconds [26] for the DC power to change from 0 to the rated value, while the ramping time of generators is several hours [27]. Compared to generators, the power regulation time of HVDC can be ignored in the restoration process. Thus, there may be some loop paths with better restoration efficiency that will be restored earlier than generators with low efficiencies (long start-up times or slow ramping rates). The restoration of these loop paths can effectively increase the system strength and support a better restoration of LCC-HVDC. As a result, the loop structure will appear in the restoration paths forming a non-tree skeleton restoration network.

In summary, when making the restoration path scheme for the AC-DC power grid, it should consider both the rapidity of grid restoration and the improvement of system strength to make full use of LCC-HVDC and improve the restoration efficiency. Therefore, this paper proposes a novel method of restoration path optimization for the AC-DC bulk power grid. The main contributions are as follows:
1) The supporting capability of the AC power grid for the LCC-HVDC in the restoration process is analysed from the aspects of start-up and operation of LCC-HVDCs based on the system strength.
2) The quantitative relationship between the restoration path and the restoration characteristic of LCC-HVDC is derived and a new function of the restoration path that enhances the system strength is proposed.
3) The concept of a "non-tree restoration path" is introduced. Then an optimization model of the restoration path that considers loop networks is formulated and a feasible optimization algorithm is proposed to solve this model.

The remainder of this paper is organized as follows. Section II analyzes the support capability of the AC power grid to the LCC-HVDC in the restoration process. Section III quantifies the relationship between the restoration path and the LCC-HVDC restoration characteristic. Section IV proposes the restoration path optimization model and an optimization algorithm considering non-tree restoration paths. In Section V, case studies are conducted to verify the effectiveness of the proposed method. Finally, conclusions are presented in Section VI.

## 2. Analysis of the AC Power Grid Supporting the LCC-HVDC in the Restoration Process

Regarded as external sources, LCC-HVDCs can provide support power for blackout power grids and speed up the restoration process, while LCC-HVDCs must get enough support to restart and operate from AC power grids with sufficient system strength. Thus, this study analyzes the support capability of AC power grids to LCC-HVDCs.

### 2.1. AC System Strength for Supporting the LCC-HVDC Start-up

There are requirements for the minimum continuous direct current and the minimum number of filters when the LCC-HVDC starts up [21]. The real minimum continuous current is for avoiding DC intermittence, which is 5–10% of the rated value. The minimum number of filters is used to filter out harmonics of different frequencies separately. Thus, the start-up of LCC-HVDC will bring impulses of active power and reactive power to the restored AC power grid. These impulses lead to big changes in frequency and voltage. If the changes in voltage and frequency exceed the tolerance of restored power grids, the grid may experience a blackout again. Therefore, the system strength of the restored power grid must be sufficient to prevent the voltage and frequency from exceeding their limits.

This study introduces two system strength indices [24] to reflect the support capability of the restored AC power grid, which are the short circuit capacity (SCC) of node $d$ where an LCC-HVDC locates and the frequency regulation capability (FRC) of the restored grid. The SCC refers to the apparent power when a three-phase short circuit event occurs. It reflects the strength of the power system's supply capability. The FRC pertains to a system's capability to maintain the frequency within acceptable limits. It is mainly determined by the operating generators, especially during the power grid restoration process. They are respectively expressed as:

$$\begin{cases} S_{sc,d}(t) = \dfrac{U_D^2(t)}{z_{dd}(t)} \\ M_f(t) = \sum_{g=1}^{n_g(t)} \delta_g(t) \dfrac{P_{GN,g}}{df_g} \end{cases} \quad (1)$$

where $z_{dd}(t)$ and $n_g(t)$ are determined by the restored AC power grid. $df_g$ can be calculated based on [28]. If $\delta_g(t)$ is 1, it means the generator $g$ has connected to the power grid; otherwise, $\delta_g(t)$ is 0.

Given the relatively weak nature of restored power grids in the early restoration process, the LCC-HVDC should employ a suitable start-up mode and control strategy to minimize its impact on restored grids. Generally, the LCC-HVDC uses the minimum single-pole mode to minimize the impact and ensure a successful start-up. The DC voltage is 70% of the rated voltage, the current is 10% of the rated current [20]. The control mode is the constant current at the rectifier and the constant voltage at the inverter mode [20]. The filter with minimum capacity is chosen to prevent overvoltage caused by excessive reactive power injection. Thus, the lower limits of SCC and FRC are:

$$\begin{cases} S_{sc}^{min} = \dfrac{U_D(t)}{\Delta U_{max}}[Q_F - Q_D(t)] \\ M_f^{min} = \dfrac{0.7 \times 0.1}{2\Delta f_{max}} P_{DN} \end{cases} \quad (2)$$

where $\Delta U_{max}$ and $\Delta f_{max}$ are set as $0.1U_N$ and 0.5Hz, respectively. Based on (1)-(2), the support capability constraints of the restored AC power grid can further be derived. The constraints of the SCC and FRC are expressed as:

$$\begin{cases} S_{sc,d}(t) \geq S_{sc}^{min} \\ M_f(t) \geq M_f^{min} \end{cases} \quad (3)$$

where $S_{SC,d}(t)$ and $M_f(t)$ both constantly change with the restoration of AC power grids.

### 2.2. AC System Strength for Supporting the LCC-HVDC Operation

The system strength of the restored AC power grid is still relatively weak after the LCC-HVDC start-up. To ensure the restored grid operates stably, the LCC-HVDC should remain in its start-up mode unchanged until the skeleton network is reconfigured and all generators are restarted.

When the LCC-HVDC is operating, the system strength mainly depends on the relative capacity of the restored AC grid and the LCC-HVDC system, which is presented as the short circuit ratio (SCR) [29]. SCR is defined as the ratio of the SCC to the LCC-HVDC transmission power. It can effectively reflect the support capability of the AC power grid for the LCC-HVDC. The SCR is expressed as:

$$S_{CR,d}(t) = \dfrac{S_{sc,d}(t)}{P_D(t)} \quad (4)$$

According to the categorization criteria of system strength [29], the minimum SCR for the restored AC grid is set as 3 in this study, which is sufficient for supporting the stable operation of the LCC-HVDC. Based on (4), it can be seen that when the minimum value of $S_{CR}$ is set as 3, the upper limit of $P_D$ is proportional to $S_{sc}$, which is expressed as:

$$P_D^{up}(t) = \frac{1}{3} S_{sc,d}(t) \quad (5)$$

Eq. (5) shows that at restoration time $t$, the upper limit of the LCC-HVDC's transmission power is determined by the restored AC power grid. Meanwhile, with the continued restoration, the system strength of restored AC power grids and the upper limit of $P_D$ continue increasing. In addition, the output power of LCC-HVDC can be gradually increased by adjusting the predetermined value of controllable direct current.

## 3. Quantitative Relationship Between Restoration Paths and the Restoration Characteristic of LCC-HVDC

The system strength is the bridge between the restoration path and the LCC-HVDC restoration characteristic. At restoration time t, the restored paths represent the restored AC power grid and determine the present system strength. Meanwhile, the system strength determines the state of LCC-HVDC start-up and operation. Therefore, there is a quantitative relationship between the restoration path and LCC-HVDC restoration characteristic. Based on equation (1), the $P_{GN,g}/df_g$ and $z_{dd}$ are used as indices for FRC and SCC to quantify the relationship, respectively.

### 3.1. Influence of the Restoration Path on the FRC

Since $P_{GN,g}/df_g$ is a generator's characteristic, the FRC is only determined by generators that have operated in the restored grid. That is, FRC will only change if a generator has restarted and is connecting to the restored grid.

Based on (1), it can be seen that $M_f$ is directly proportional to $P_{GN,g}$ and inversely proportional to $df_g$. At restoration time $t$, the operating generators are known, so $M_f$ can be calculated. With the continued power grid restoration, more generators are started, and it is easier to reach the minimum start-up limit, $M_f^{min}$.

The path restoration sequence determines the generators' start-up sequence, so as to determine which generators can participate in frequency regulation at different restoration times. Thus, the restoration path indirectly impacts the FRC of restored systems. By optimizing the restoration path, the FRC of the restored AC grid can be improved, and the speed of LCC-HVDC start-up can be accelerated.

### 3.2. Influence of the Restoration Path on the SCC

The SCC is related to the whole restored AC grid and $z_{dd}$ is calculated based on the parameters and topology of the restored grid. As $z_{dd}$ includes all the information of the restored grid, any change in the restoration path will change the value of $z_{dd}$.

According to (1), $z_{dd}$ is inversely proportional to the $S_{SC,d}$. In this study, the Thevenin equivalent impedance is used to analyze the influence of the restoration path on the SCC. We introduce the branch-adding method [30] to form the nodal impedance matrix $\mathbf{Z}$, which can analogize the restoration process based on the path restoration sequence.

Based on the graph theory, each electrical component in the power grid is regarded as a branch. Without considering the non-linear electrical component, we set the restored new branch as $\alpha$, the nodes at both ends of the branch $\alpha$ are $p$ and $q$. Node $p$ is already restored and node $q$ is going to be restored.

*3.2.1 Influence of the Tree-branch:* In the power grid, the tree-branch generally corresponds to transmission lines and transformers. The influence of tree-branches on $z_{dd}$ can be quantified based on the equation of adding tree-branches. When restoring a tree-branch $\alpha$, if the restored grid contains node $d$, $z_{dd}$ and $S_{SC,d}$ both remain unchanged. Otherwise, $q$ equals $d$, and $z_{dd}$ is:

$$z_{dd} = z_{qq} = z_{\alpha\alpha} + z_{pp} \quad (6)$$

where $z_{dd}$ can be obtained by adding the self-impedance of connected nodes and the tree-branch impedance between them. By analogy, the influence of all tree-branches on the SCC in the restoration path can be quantified. In the restoration process, the restoration of tree-branches corresponds to the charging operation of transmission lines and transformers.

*3.2.2 Influence of the Ungrounded Link-branch:* The ungrounded link-branch generally corresponds to loop paths, which include transmission lines and transformers. The restoration path will be a non-tree structure when loop paths appear. Based on the equation of adding link-branches, when restoring an ungrounded link-branch $\alpha$, if the restored grid contains node $d$, $z_{dd}$ will change to:

$$z'_{dd} = z_{dd} - \frac{(z_{dp} - z_{dq})^2}{z_{\alpha\alpha} + z_{pp} + z_{qq} - z_{pq} - z_{qp}} \quad (7)$$

Since $z_{pp}$ is greater than $z_{pq}$, and $z_{qq}$ is greater than $z_{qp}$, $z_{dd}$ will decrease and $S_{SC,d}$ will increase after restoring ungrounded link-branches. If the restored grid does not contain node $d$, the self-impedance of the previous sequence nodes related to $z_{dd}$ will decrease and further affect $z_{dd}$. By analogy, the influence of all ungrounded link-branches on the SCC can be quantified.

In the restoration process, the restoration of ungrounded link-branches corresponds to the loop-closing operation. The link-branch forms a loop path with related tree-branches restored in the previous sequence. Therefore, branches contained in a loop path can be regarded as a generalized ungrounded link-branch.

*3.2.3 Influence of the Grounded Link-branch:* The grounded link-branch generally corresponds to generators. Its influence on $z_{dd}$ can be quantified based on the equation of adding link-branches. When restoring a grounded link-branch $\alpha$, if a restored grid contains node $d$, $z_{dd}$ changes to:

$$z'_{dd}(t) = z_{dd}(t) - \frac{z_{dp}^2}{z_{\alpha\alpha} + z_{pp}} \quad (8)$$

Since $z_{dp}$, $z_{\alpha\alpha}$, and $z_{pp}$ are all bigger than 0, $z_{dd}$ will decrease and $S_{SC,d}$ will increase after restoring grounded link-branches. If the restored grid does not contain node $d$, the self-impedance of the previous sequence nodes related to $z_{dd}$ will decrease and further affect $z_{dd}$. By analogy, the influence of all grounded link-branches on the SCC can be quantified. In the restoration process, the restoration of grounded link-branches corresponds to the grid-connection operation of generators.

In addition, the impedance of the load and parallel reactive equipment is much greater than the impedance of generators, transmission lines and transformers, so their influences on the SCC are ignored in this study.

In summary, with the continued restoration of the power grid, FRC and SCC continue increasing. By optimizing the restoration path, $P_{GN,g}/df_g$ and $z_{dd}$ can be improved and the system strength can be enhanced effectively. Thus, the start-up time of LCC-HVDC can be shortened and the upper limit of transmission power of LCC-HVDC can be raised. Then, the black-out power grid can get more support power from LCC-HVDC for restoration.

### 3.3. Restoration Path-based LCC-HVDC Restoration Characteristic

The restoration characteristic of LCC-HVDC is significantly different from the ramping characteristic of generators. Compared to generators, the ramping time of LCC-HVDC can be ignored in the restoration process. Hence, the LCC-HVDC can improve the restoration speed of the power grid. However, the transmission power of LCC-HVDC is finite, which depends on the system strength of AC power grids connected to both ends of the LCC-HVDC system. This study focuses on the power grid restoration at the inverter side and assumes that the power grid at the rectifier side operates normally and is strong enough.

Based on (5), it can be seen that the upper limit of LCC-HVDC transmission power is proportional to the SCC when the lower limit of SCR is given. That is, $P_D^{up}$ is determined by $S_{SC,d}$. Affected by the restoration path sequence, $P_D^{up}$ is:

$$P_D^{up}(t) = \begin{cases} 0, & 0 \leq t < t_{start} \\ P_{D,1}^{up}, & t_{start} \leq t < t_1 \\ \cdots & \cdots \\ P_{D,m}^{up}, & t_m \leq t < t_{m+1} \\ \cdots & \cdots \\ P_{D,end}^{up}, & t \geq t_{end} \end{cases} \quad (9)$$

As shown in Fig. 1, $P_D^{up}$ shows a step-by-step uptrend with the continued restoration. It should be noted that Fig. 1 serves as a conceptual portrayal and illustrates the characteristic of step-wise increase for $P_D^{up}$. While the precise values on the curve may vary for different restoration paths, the step-like rising trend of the curve remains consistent.

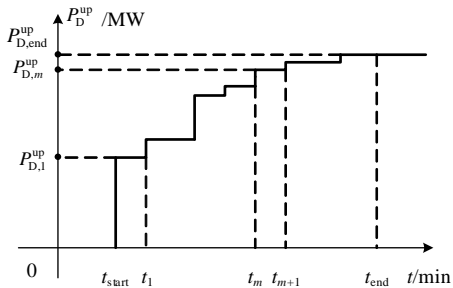

Fig. 1. The schematic curve of characteristic for LCC-HVDC power.

To sum up, the main factors affecting the LCC-HVDC restoration characteristic are the parameters and restored sequence of restoration paths. Therefore, by optimizing the restoration path, the LCC-HVDC restoration characteristic can be improved. With the full usage of LCC-HVDC power, the power grid restoration speed can be accelerated.

## 4. Model and Solution of the restoration path optimization for AC-DC power grids

### 4.1. Restoration Path Considering the Non-Tree Path

Based on the quantitative analysis in Section II, it can be seen that different restoration paths and their sequences have different impacts on LCC-HVDC for supporting the power grid restoration. In particular, link-branches have significant impacts on the system strength. Thus, we propose the concept of the "non-tree path" in this study.

Relative to the concept of "tree" in the graph theory, "non-tree" refers to a graph with loops, *i.e.*, there are closed loops between some node pairs in the graph. Corresponding to the power grid, the "non-tree path" is a branch set that can form a loop network.

Considering the LCC-HVDC restoration characteristic, the traditional optimal tree path may not be suitable for the restoration of the AC-DC power grid. Sometimes, non-tree paths near LCC-HVDCs (electrical distance) may have better restoration efficiency in the restoration process. Therefore, the non-tree path is of great necessity when establishing the optimization model of the restoration path.

### 4.2. Objective Function

The ultimate goal of the power grid restoration is to restore all loads as soon as possible. In the early restoration stage, the system restoration speed and restored load are limited by restarted power sources. Thus, this study takes the minimum negative ratio of the total supplied power to restoration time as the objective function. The objective function can effectively reflect the restoration efficiency.

$$\min: F = -[\sum_{g=1}^{N_G} \int_0^T \delta_g(t) P_{G,g}(t) dt + \int_0^T P_D(t) dt]/T \quad (10)$$

In the restoration process, once a generator is started, its output increases according to the ramping rate until reaching the rated power. According to [11], the simplified equation of $P_{G,g}(t)$ is expressed as:

$$P_{G,g}(t) = \begin{cases} 0, & 0 \leq t < t_{s,g} \\ r_g(t - t'_{s,g}), & t'_{s,g} \leq t < t''_{s,g} \\ P_{GN,g}, & t \geq t''_{s,g} \end{cases} \quad (11)$$

### 4.3. Constraints

The proposed restoration path optimization model should meet the following equality and inequality constraints.

$$\begin{cases} z_{dd}(t) \leq \dfrac{U_D(t)\Delta U_{max}}{2Q_F - Q_D(t)} \\ \sum_{g=1}^{N_G} \delta_g(t)(P_{GN,g}/df_g) \geq \dfrac{0.035 P_{DN}}{\Delta f_{max}} \end{cases} \quad (12)$$

$$P_D(t) = N_r[1.35 E_{LL}(t)\cos\gamma(t) I_D(t) - \dfrac{3}{\pi} X_r I_D^2(t)] \quad (13)$$

$$Q_{\mathrm{D}}(t) = P_{\mathrm{D}}(t)\sqrt{(U_{\mathrm{D0}}/U_{\mathrm{D}}(t))^2 - 1} \quad (14)$$

$$P_{\mathrm{D}}(t) \leq \frac{S_{\mathrm{sc}}(t)}{3} \quad (15)$$

$$P_{\mathrm{D}}^{\min} \leq P_{\mathrm{D}}(t) \leq \min\{P_{\mathrm{D}}^{\mathrm{up}}(t), P_{\mathrm{DN}}\} \quad (16)$$

$$P_{\mathrm{G},g}^{\min} \leq P_{\mathrm{G},g}(t) \leq P_{\mathrm{G},g}^{\max} \quad (17)$$

$$Q_{\mathrm{G},g}^{\min} \leq Q_{\mathrm{G},g}(t) \leq Q_{\mathrm{G},g}^{\max} \quad (18)$$

$$\begin{cases} 0 < t_{s,g} \leq T_{\mathrm{CH},g} \\ t_{s,g} \geq T_{\mathrm{CC},g} \end{cases} \quad (19)$$

$$P_{\mathrm{D}}(t) - P_{\mathrm{DL}} + \sum_{g=1}^{N_{\mathrm{G}}}\{\delta_g(t)[P_{\mathrm{G},g}(t) - P_{\mathrm{cr},g}]\} \geq \sum_{i=1}^{N_{\mathrm{B}}}\delta_i(t)P_{\mathrm{B},i}(t) \quad (20)$$

$$\sum_{l=1}^{N_{\mathrm{L}}} \delta_l(t) Q_{\mathrm{L},l} - \sum_{i=1}^{N_{\mathrm{B}}} \delta_i(t) Q_{\mathrm{B},i}(t) + Q_{\mathrm{filter}}(t) - Q_{\mathrm{D}}(t) \\ \leq \min\{\sum_{g=1}^{N_{\mathrm{G}}} Q_{\mathrm{A},g}^{\max}, \sum_{g=1}^{N_{\mathrm{G}}} K_{\mathrm{CB},g} S_{\mathrm{GN},g}\} \quad (21)$$

$$P_{\mathrm{G},i}(t) + P_{\mathrm{D},i}(t) - P_{\mathrm{B},i}(t) \\ = U_{\mathrm{B},i}(t)\sum_{j=1}^{N_{\mathrm{B}}} U_{\mathrm{B},j}(t)(G_{ij}\cos\theta_{ij} + B_{ij}\sin\theta_{ij}) \quad (22)$$

$$Q_{\mathrm{G},i}(t) + [Q_{\mathrm{D}}(t) - Q_{\mathrm{filter}}(t)] - Q_{\mathrm{B},i}(t) \\ = U_{\mathrm{B},i}(t)\sum_{j=1}^{N_{\mathrm{B}}} U_{\mathrm{B},j}(t)(G_{ij}\sin\theta_{ij} - B_{ij}\cos\theta_{ij}) \quad (23)$$

$$P_{\mathrm{L},l}^{\min} \leq P_{\mathrm{L},l}(t) \leq P_{\mathrm{L},l}^{\max} \quad (24)$$

$$U_{\mathrm{B},i}^{\min} \leq U_{\mathrm{B},i}(t) \leq U_{\mathrm{B},i}^{\max} \quad (25)$$

where $N_r$ equals 1, $P_{\mathrm{D}}^{\min}$ equals 0.035 $P_{\mathrm{DN}}$, $P_{\mathrm{DL}}$ is 0.01*$P_{\mathrm{DN}}$.

Constraints (12)-(16) impose the LCC-HVDC-related constraints. Constraint (12) denotes the start-up limits of the restored AC power grid for the LCC-HVDC. Equations (13)-(16) denote the active and reactive power of the LCC-HVDC inverter exchanged with the AC grid [31]. Constraint (15) gives the upper limit for DC active power based on the restored AC grid. Constraint (16) is the lower and upper limits for the DC active power of the LCC-HVDC. In addition, we keep $S_{\mathrm{CR}}(t) \geq 3$ in the whole restoration process to ensure the stable operation of LCC-HVDC. Constraints (17)-(19) explain the generator-related constraints. Constraints (17) and (18) show the lower and upper limits of active and reactive power for generators. Constraint (19) denotes that the restarted thermal generators must meet the time limit of hot start-up and cold start-up. Constraints (20)-(25) denote the power flow-related constraint. Constraint (20) ensures that the restarted sources can supply enough active power for restoring off-line generators and load. Constraint (21) denotes the reactive power limit in the restored power grid. Equations (22) and (23) denote the AC-DC combined power flow which is solved by the alternating iteration method [32]. Constraint (24) denotes the capacity limit of the power flow in branches [33]. Constraint (25) denotes the lower and upper limits of bus voltage[34].

*4.4. Optimization Model Solving*

As the restored paths may form a non-tree network including loops, this study proposes an improved topological sorting and multi-population genetic algorithm (ITS-MPGA) to solve the optimal restoration path scheme. Based on the MPGA algorithm [35], the ITS algorithm [36] is introduced to generate the feasible path restoration sequence. The proposed algorithm can achieve the optimal solution by only searching in the discrete feasible solution space, with the characteristic of a small searching space and high searching efficiency.

The ITS-MPGA algorithm includes two submodules: the path restoration scheme (PRS) generation module based on the ITS algorithm and the optimal PRS solution module based on the MPGA algorithm. In the algorithm, the chromosome is formed in arranged numbers, each gene in chromosomes represents a branch number and each chromosome represents a path restoration scheme. The traditional topology sorting algorithm is improved by using the priority weight. The MPGA algorithm is modified to integer encoding.

In the PRS generation module, a set of PRS is randomly generated first. Second, based on the priority relationship of restoration paths, the path restoration sequence is modified to meet the network connectivity constraint. Then, the feasible PRS is obtained by using the ITS algorithm.

In the optimal PRS solution module, first, the objective function value of feasible PRS is calculated, the constraints are checked, and those schemes that do not meet constraints are deleted. Second, a new feasible PRS set is formed by the operation of adaptive fragment crossover, mutation and amendment, with calculating objective function values and checking constraints. Then, the next generation of the PRS set is obtained by using the operations of elite selection and migration. By analogy, the algorithm will end when the maximum evolutionary algebra is reached. Finally, the optimal scheme is obtained.

The solution flow chart of the restoration path optimization model based on the ITS-MPGA algorithm is shown in Fig. 2.

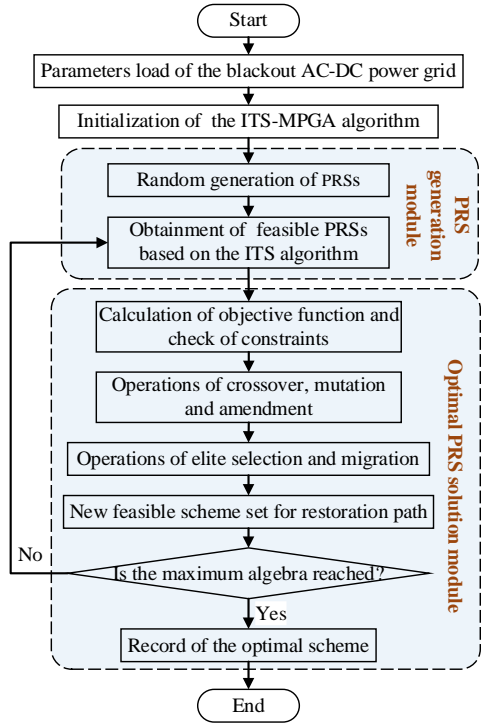

Fig. 2. The solution flow chart of the restoration path optimization model.

## 5. Case Studies

### 5.1. The Modified IEEE 39-Bus System

*5.1.1 Set up:* In the modified IEEE 39-bus system, the LCC-HVDC locates at node 39, *i.e.*, $d=39$, which replaces the initial generator. Assuming $P_{DN}$ is set as 1000 MW, $O_F$ in the inverter station is 120Mvar. Then, $S_{sc}^{min}$ and $M_f^{min}$ are calculated as 850 MVA and 70 MW/Hz based on (2), respectively. The time consumption of LCC-HVDC from charging to grid-connection is 10 minutes.

The black start power source is set as the generator on node 31. The restoration-related parameters of each generator are shown in Table 1. The restoration time consumption of transmission lines and transformers are both set as 5 minutes, and the calculation time step in the restoration process is set as 5 minutes as well.

In the ITS-MPGA algorithm, the maximum evolutionary algebra is 200, the migration rate is 0.2, the number of sub-populations is 6, and the size of each subpopulation is 10.

Table 1. Restoration-related parameters of generators in the modified IEEE 39-bus system

| Generator node | $P_{GN,g}$(MW) | $x'_g$ (p.u.) | $df_g$ (Hz/p.u.) | $r_g$ (MW/h) | Time consumption of grid-connection (min) |
|---|---|---|---|---|---|
| 30 | 250 | 0.031 | 3.40 | 150.0 | 10 |
| 31 | 600 | 0.070 | 7.94 | 360.0 | 0 |
| 32 | 650 | 0.053 | 7.94 | 390.0 | 30 |
| 33 | 632 | 0.044 | 4.61 | 379.2 | 30 |
| 34 | 508 | 0.132 | 4.61 | 304.8 | 30 |
| 35 | 650 | 0.050 | 7.94 | 390.0 | 30 |
| 36 | 560 | 0.049 | 4.61 | 336.0 | 30 |
| 37 | 540 | 0.057 | 7.94 | 324.0 | 30 |
| 38 | 830 | 0.057 | 7.94 | 498.0 | 30 |

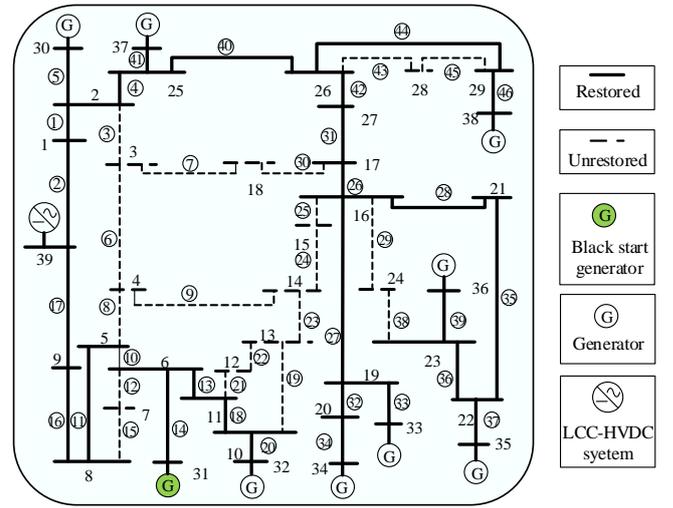

Fig. 3. The final skeleton network of Scheme 1.

*5.1.2 Result analysis:* The optimal restoration path scheme (Scheme 1) based on the proposed method is shown in Fig. 3 and Table 2. Fig. 3 shows the final skeleton network of Scheme 1. Table 2 shows the restoration process by giving a detailed path restoration sequence. A restoration stage means there is a generator that connects to the grid or a loop path appears.

The $T$ is 370 minutes, which equals the last power source connected to the restored grid. The $F$ value is -2107.19 MW. The LCC-HVDC starts up at the 105th minute, the $S_{sc}$ of the restored AC power grid is 1700.97 MVA and the $M_f$ is 331.86MW/Hz at this time. The start-up of LCC-HVDC causes voltage deviation, $\Delta U$, of node 39 and frequency deviation, $\Delta f$, of the grid are 0.050 p.u. and 0.105 Hz, respectively. The $P_D(t)$ curve of LCC-HVDC is shown with the blue solid line in Fig. 4. In the restoration process, the $P_D$ is shown in Table 2 at each generator's grid-connection time. Finally, $P_D$ reaches 645.95 MW.

In addition, although node 39 is charged at the 70th minute, the LCC-HVDC does not start up. This is because the $S_{sc}$ is only 832.92 MVA at this time, which is less than the lower limit. The restored AC grid is not strong enough to support the safe start-up of LCC-HVDC.

The results show that the support effect of the restoration path on the LCC-HVDC is inversely proportional to the electrical distance between them. Namely, restoration paths adjacent to the LCC-HVDC have more support effects, especially generator paths or loop paths. It is very beneficial to quickly reach the LCC-HVDC start-up limits and improve the upper limit of DC transmission power if their electrical distance is small. On the contrary, the greater the electrical distance is, the smaller the impact of paths on the LCC-HVDC is.

Table 2. Optimal scheme of restoration path (Scheme 1)

| Stage | Source node | Start-up time(min) | Grid-connection time(min) | Path restoration sequence | $P_D$ (MW) |
|---|---|---|---|---|---|
| 1 | 31 | 0 | 0 | -- | 0 |
| 2 | 32 | 20 | 50 | ⑭-⑬-⑱-⑳ | 0 |
| 3 | 39 | 70 | 105 | ⑩-⑪-⑯-⑰ | 0 |
| 4 | 30 | 95 | 105 | ②-①-⑤ | 566.99 |
| 5 | 37 | 115 | 145 | ④-㊶ | 618.76 |
| 6 | 38 | 160 | 190 | ㊵-㊹-㊻ | 633.56 |
| 7 | 33 | 215 | 245 | ㊷-㊱-㉖-㉗-㊳ | 642.37 |

| | | | | | | |
|---|---|---|---|---|---|---|
| 8 | 35 | 260 | 290 | ㉘-㉟-㊲ | | 644.90 |
| 9 | 36 | 300 | 330 | ㊱-㊴ | | 645.57 |
| 10 | 34 | 340 | 370 | ㉜-㉞ | | 645.95 |

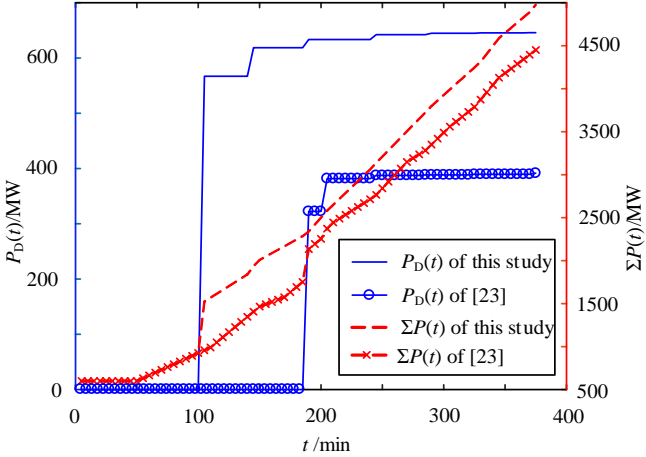

Fig. 4. The $P_D(t)$ curve and $\sum P(t)$ curve for the LCC-HVDC.

*5.1.3 Comparative analysis with traditional optimal tree path restoration scheme:* To illustrate the effectiveness of the proposed method considering the non-tree path, we compare the optimal restoration path schemes based on the proposed method and [23]'s shortest path method (Scheme 2).

In Scheme 2, the power sources' start-up sequence is 31-32-37-35-39-30-33-34-36-38. The connection paths between sources are searched based on the Dijkstra algorithm, and then the path restoration sequence and the final skeleton network are obtained. The transmission power when the LCC-HVDC is operating is calculated based on (4) and (9), as [23] does not give the operation characteristic of LCC-HVDC.

For Scheme 2, the $T$ is 375 minutes and $F$ is -1999.27 MW. The LCC-HVDC starts up at the 190th minute, the $S_{sc}$ of the restored AC power grid is 1026.8 MVA and the $M_f$ is 250.00 MW/Hz. The $\Delta U$ and $\Delta f$ of node 39 caused by the start-up of LCC-HVDC are 0.086 p.u. and 0.140 Hz, respectively. The $P_D(t)$ curve is shown with the blue dotted line in Fig. 4. Finally, the $P_D$ reaches 391.67 MW. Fig. 4 compares the curves of $P_D(t)$ and the total power output, $\sum P(t)$, of Scheme 1 and Scheme 2. The differences between the two schemes are as follows:

1) Compared with Scheme 2, Scheme 1 supports the LCC-HVDC start-up earlier. The $\Delta U$ and $\Delta f$ of Scheme 1 are both smaller, which means that the system strength of the restored grid is higher when restarting the LCC-HVDC. The LCC-HVDC can start up more safely.
2) Scheme 1 can support the LCC-HVDC to transmit more DC power than Scheme 2 at any restoration moment. Moreover, the final $P_D$ value of Scheme 1 is 64.92% higher than Scheme 2. The LCC-HVDC's performance of Scheme 1 is better in supporting the power grid restoration.
3) The total restoration time, $T$, of Scheme 1 is smaller. After 105 minutes, the $\sum P(t)$ of Scheme 1 keeps bigger than Scheme 2. These show that Scheme 1 can restore more load in a shorter time, it improves the restoration efficiency and speeds up the restoration process.

## 5.2. A Partial Power Grid of Southwest China

The partial power grid of Southwest China contains 51 nodes and 60 branches. The restoration-related parameters of all seven generators are shown in Table 3. The black start generator is "WM". The name of the LCC-HVDC is "RY", which connects to the "HR" station. The $P_{DN}$ is 3000 MW and $Q_F$ is 160 MVar. So the $S_{sc}^{min}$ and $M_f^{min}$ are 2150 MVA and 210 MW/Hz, respectively.

The optimal path restoration scheme (Scheme 3) based on the proposed method is shown in Table 4. The $T$ is 295 minutes and the $F$ is -2179.84 MW. The LCC-HVDC starts up at the 110th minute, and the $S_{sc}$ and $M_f$ of the restored AC grid at this moment are 2243.74 MVA and 438.98 MW/Hz, respectively. The $\Delta U$ and $\Delta f$ of node 39 caused by the start-up of LCC-HVDC are 0.096 p.u. and 0.239 Hz. The curves of $P_D(t)$ and $\sum P(t)$ are shown in Fig. 5. The final $P_D$ reaches 1451.70 MW. Table 4 shows the restoration process of Scheme 3 in detail.

In particular, the restored paths form a non-tree network with three loops as shown in Fig. 6 (a). The path sets of the three loops correspond to the "path restoration sequence" column of restoration stages 7, 8 and 10. In the restoration process, paths of stages 7 and 8 are restored before generator "SJ" and "HEX", and paths of stage 10 are restored before generator "HEX". That is because the three loop paths increase the DC transmission power, which has a better restoration effect than subsequent generators. Meanwhile, it can be seen that the closer the loop path is to the LCC-HVDC, the greater the support effect is on the LCC-HVDC.

Based on the method of [23], the power sources' start-up sequence is WM-MK-QGG-RY-YJ-YL-SJ-HEX. The corresponding final skeleton network is given in Fig. 6 (b), which is a tree network. Then the optimal path restoration scheme (Scheme 4) can be found based on Fig. 6 (b). In Scheme 4, the $T$ is 255 minutes and $F$ is -1715.70 MW. The LCC-HVDC starts up at the 110th minute, and the $S_{sc}$, $M_f$, $\Delta U$ and $\Delta f$ are the same as Scheme 3 at this time. The curves of $P_D(t)$ and $\sum P(t)$ are shown in Fig. 5. The final $P_D$ is 869.76 MW.

By comparing Scheme 3 and Scheme 4, it can be seen that: 1) Although the restoration time of Scheme 3 is longer, its $P_D(t)$ and $\sum P(t)$ are both larger than Scheme 4 after the LCC-HVDC start-up, and the value of the objective function is smaller. This indicates that Scheme 3 can restore more loads at the same time and has higher restoration efficiency. 2) When the LCC-HVDC is operating, Scheme 3 can promote the system strength faster. The final $P_D$ value of Scheme 3 is 66.91% higher than Scheme 4. Therefore, the LCC-HVDC can provide more support power for the restoration of the AC-DC power grid.

Table 3. Restoration-related parameters of generators in the partial power grid of Southwest China

| Generator node | $P_{GN,g}$(MW) | $x'_g$ (p.u.) | $df_g$ (Hz/p.u.) | $r_g$ (MW/h) | Time consumption of grid-connection (min) |
|---|---|---|---|---|---|
| SJ | 180 | 0.086 | 3.40 | 108 | 10 |
| QGG | 450 | 0.040 | 3.40 | 270 | 10 |
| YJ | 600 | 0.066 | 3.40 | 360 | 10 |
| WM | 600 | 0.031 | 3.40 | 360 | 0 |
| YL | 600 | 0.036 | 3.40 | 360 | 10 |
| MK | 600 | 0.024 | 4.61 | 360 | 30 |
| HEX | 140 | 0.131 | 3.40 | 84 | 10 |

Table 4. Optimal scheme of restoration path (Scheme 3)

| stage | Source node | Start-up time (min) | Grid-connection time (min) | Path restoration sequence | $P_D$ (MW) |
|---|---|---|---|---|---|
| 1 | WM | 0 | 0 | -- | 0 |
| 2 | MK | 25 | 55 | ㊽-②-③-㊽-㊼ | 0 |
| 3 | QGG | 75 | 85 | ①-④-⑮-㊵ | 0 |
| 4 | RY | 100 | 110 | ⑤-⑰-㉓ | 747.91 |
| 5 | YL | 135 | 145 | ㉒-㉕-㉘-㉖-㊴ | 1135.48 |
| 6 | YJ | 170 | 180 | ⑱-㊷-㊶-㊸-㊹ | 1196.67 |
| 7 | -- | -- | 185 | ⑭ | 1317.71 |
| 8 | -- | -- | 200 | ㊻-㊾-㊿ | 1378.95 |
| 9 | SJ | 230 | 240 | ㉗-㉚-㉜-㉛-㉝-㉟ | 1409.11 |
| 10 | -- | -- | 260 | ⑬-⑲-㊹-㉔ | 1437.87 |
| 11 | HEX | 285 | 295 | ⑯-⑫-⑪-㊻-㊽ | 1451.70 |

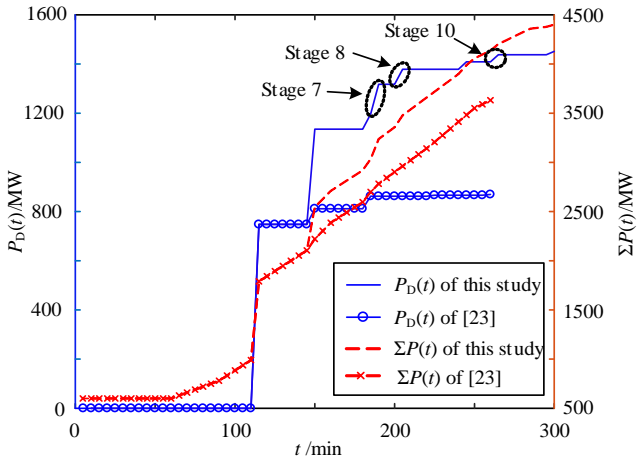

Fig. 5. The $P_D(t)$ curve and $\sum P(t)$ curve for the LCC-HVDC.

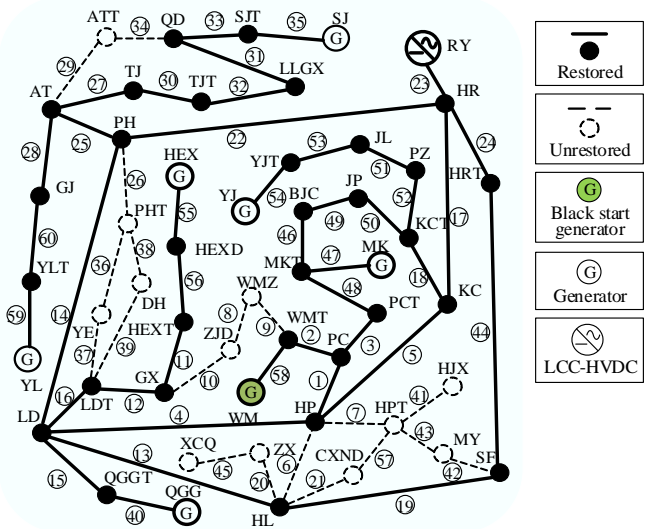

Fig. 6. (a) The final skeleton network of Scheme 3.

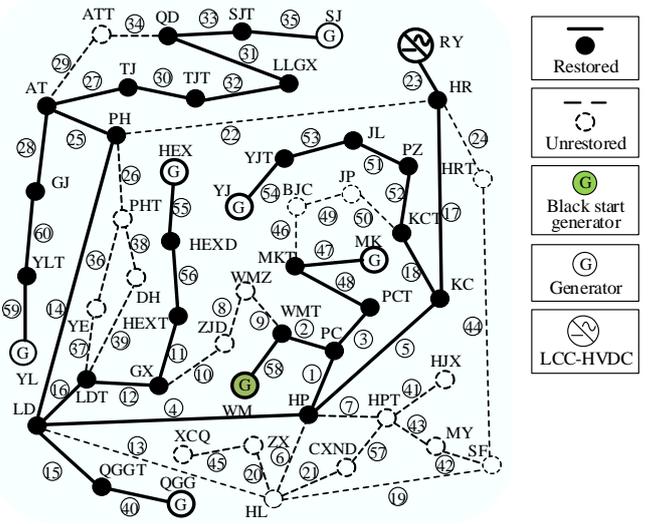

Fig. 6. (b) The final skeleton network of Scheme 4.

## 6. Conclusion

This paper proposes a novel method of restoration path optimization for the AC-DC bulk power grid after a major blackout. The proposed method can effectively utilize the power support capability of LCC-HVDCs and improve the restoration efficiency of AC-DC power grids. Simulation results show that: 1) The participation of LCC-HVDC can speed up the grid restoration. 2) Only by improving the system strength can the support capability of LCC-HVDC be effectively used. 3) The restoration path and its sequence determine the system strength and the LCC-HVDC restoration characteristics. 4) The restoration strategy of the AC-DC power grid must take into account the non-tree path to further improve the restoration efficiency.

In addition, this study does not consider the voltage source converter-based high voltage direct current (VSC-HVDC) system and modular multilevel converter-based high voltage direct current (MMC-HVDC) system. Further research will focus on the restoration strategy for the AC-DC power grid with the hybrid multi-HVDC system.

## 7. Funding

This work is supported by the National Natural Science Foundation of China under Grant 52107092.

### References

[1] Wang P., Goel L., Liu X., Choo F. H.. Harmonizing AC and DC: a hybrid AC/DC future grid solution. *IEEE Power and Energy Magazine*, 11(3) (2013), pp. 76-83.
[2] Li, Y., Wang, R., Li, Y., Zhang, M., & Long, C.. Wind power forecasting considering data privacy protection: A federated deep reinforcement learning approach. Applied Energy, 329 (2023), 120291.
[3] Sun Y., Xia D., Gao Z., Wang Z., Li G., Lu W., Wu X., Li, Y.. Probabilistic load flow calculation of AC/DC hybrid system based on cumulant method. *International Journal of Electrical Power & Energy Systems*, 139 (2022), pp. 107998.
[4] Oni O. E., Davidson I. E., Mbangula K. N. I.. A review of LCC-HVDC and VSC-HVDC technologies and applications. *2016 IEEE 16th International Conference on Environment and Electrical Engineering (EEEIC)*, (2016), pp. 1-7.
[5] Yang, C., Cheng, T., Li, S., Gu, X., & Yang, L.. A novel partitioning method for the power grid restoration considering the support of

[6] Li S., Gu X., Zhou G., Li, Y.. Optimisation and comprehensive evaluation of alternative energising paths for power system restoration. *IET Generation, Transmission & Distribution*, 13(10) (2019), 1923-1932.

[7] Liu Y., Fan R., Terzija V.. Power system restoration: a literature review from 2006 to 2016. *Journal of Modern Power Systems and Clean Energy*, 4(3) (2016), pp. 332-341.

[8] Quiros-Tortos J., Terzija V.. A graph theory based new approach for power system restoration. Proc. *IEEE Grenoble PowerTech (POWERTECH)*, (2013) pp. 1-6.

[9] Liu Y., Gu X.. Skeleton-network reconfiguration based on topological characteristics of scale-free networks and discrete particle swarm optimization. *IEEE Transactions on Power Systems*, 22(3) (2007), pp. 1267-1274.

[10] Sudhakar T. D., Srinivas K. N.. Power system reconfiguration based on Prim's algorithm. *2011 1st International Conference on Electrical Energy Systems*, (2011), pp. 12-20.

[11] Sun W., Liu C., Zhang L.. Optimal generator start-up strategy for bulk power system restoration. *IEEE Transactions on Power Systems*, 26(3) (2011), pp. 1357-1366.

[12] Qiu F., Li P.. An integrated approach for power system restoration planning. *Proceedings of the IEEE*, 105(7) (July 2017), pp. 1234-1252.

[13] Ganganath N., Wang J. V., Xu X., Cheng C. T., Tse C. K.. Agglomerative clustering-based network partitioning for parallel power system restoration. *IEEE Transactions on Industrial Informatics*, 14(8) (2018), pp. 3325-3333.

[14] Li S., Lin Z., Zhang Y., Gu X., Wang H.. Optimization method of skeleton network partitioning scheme considering resilience active improvement in power system restoration after typhoon passes through. *International Journal of Electrical Power & Energy Systems*, 148 (2023), 109001.

[15] Li H., Mao W., Zhang A., Li C.. An improved distribution network reconfiguration method based on minimum spanning tree algorithm and heuristic rules. *International Journal of Electrical Power & Energy Systems*, 82 (2016), pp. 466-473.

[16] Li S., Wang L., Gu X., Zhao H., Sun Y.. Optimization of loop-network reconfiguration strategies to eliminate transmission line overloads in power system restoration process with wind power integration. *International Journal of Electrical Power & Energy Systems*, 134 (2022), pp. 107351.

[17] Ding T., Wang Z., Qu M., Wang Z., Shahidehpour M.. A sequential black-start restoration model for resilient active distribution networks. *IEEE Transactions on Power Systems*, 37(4) (2022), pp. 3133-3136.

[18] Wang T., Zhu H., Wang Z., Wang Y., Sun N., Dong Y.. Multi-objective optimization of unit restoration during network reconstruction based on DE-EDA. *2020 IEEE 3rd International Conference on Electronics and Communication Engineering (ICECE)*, (2020), pp. 102-106.

[19] Li S., Wang L., Gu X., Zhao H., Sun Y.. Optimization of loop-network reconfiguration strategies to eliminate transmission line overloads in power system restoration process with wind power integration. *International Journal of Electrical Power & Energy Systems*, 134 (2022), 107351.

[20] Li B., Liu T., Xu W., Li Q., Zhang Y., Li Y., Li X. Y.. Research on technical requirements of line-commutated converter-based high-voltage direct current participating in receiving end AC system's black start. *IET Generation, Transmission & Distribution*, 10(9) (2016), pp. 2071-2078.

[21] Li X., Liu C., Lou,Y.. Start-up and recovery method with LCC-HVDC systems participation during AC/DC system black-starts. *IET Generation, Transmission & Distribution*, 14(3) (2020), pp. 362-367.

[22] Xu K., Xie B., Wang C., Zhou Q., Xie Y., Chen X.. An ABC algorithm for optimization of restoration path in a power grid with HVDC connection. 2016 *IEEE International Conference on Smart Grid Communications (SmartGridComm)*, (2016), pp. 576-581.

[23] Li C., Xie Y., Zhou Q.. Optimization of restoration path for blackout grid with the aid of High-voltage Direct Current transmission. *Proceedings of the CSEE*, 37(9) (2017), pp. 2579-2588.

[24] Gu X., Yang C., Liang H., Li S., Zang E., Wu C.. Rapid reconfiguration of receiving-end AC local networks supporting LCC-HVDC startup for asynchronous networking power systems. *Proceedings of the CSEE*, 39(4) (2019), pp. 1060-1070.

[25] Li C., Xu Y., He J., Zhang P., Liu L.. Parallel restoration method for AC-DC hybrid power systems based on graph theory. *IEEE Access*, 7 (2019), pp. 66185-66196.

[26] Guo C., Zhang Y., Gole A. M., Zhao C.. Analysis of dual-infeed HVDC with LCC–HVDC and VSC–HVDC. *IEEE Transactions on Power Delivery*, 27(3) (2012), pp. 1529-1537.

[27] Patsakis G., Rajan D., Aravena I., Rios J., Oren S.. Optimal black start allocation for power system restoration. *IEEE Transactions on Power Systems*, 33(6) (2018), pp. 6766-6776.

[28] Adibi M. M., Borkoski J. N., Kafka R. J., Volkmann T. L.. Frequency response of prime movers during restoration. *IEEE Transactions on Power Systems*, 14(2) (1999), pp. 751-756.

[29] Zhang F., Xin H., Wu D., Wang Z., Gan D.. Assessing strength of multi-infeed LCC-HVDC systems using generalized short-circuit ratio. *IEEE Transactions on Power Systems*, 34(1) (2019), pp. 467-480.

[30] Chen M., Shi D., Li Y., Zhu L., Liu H.. Research on branches group based method for adding mutual inductance branches to Y-matrix and Z-matrix. *2014 IEEE PES General Meeting | Conference & Exposition*, (2014), pp. 1-5.

[31] Li Y., Wang H., Yang H.. A coordinated control strategy for hybrid black start with an LCC HVDC system and an auxiliary synchronous generator. *CSEE Journal of Power and Energy Systems*. (2023). (Early Access) doi: 10.17775/CSEEJPES.2021.05610.

[32] Liu C., Zhang B., Hou Y., Wu F. F., Liu Y.. An improved approach for AC-DC power flow calculation with multi-infeed DC systems. *IEEE Transactions* on *Power Systems*, 26(2) (2011), pp. 862-869.

[33] Yang C., He B., Liao H., Ruan J., Zhao, J.. Price-based low-carbon demand response considering the conduction of carbon emission costs in smart grids. *Frontiers in Energy Research*, 10 (2022), 959786.

[34] Yang C., Liu J., Liao H., Liang G., Zhao J.. An improved carbon emission flow method for the power grid with prosumers. *Energy Reports*, 9 (2023), pp. 114-121.

[35] Cochran J. K., Horng S. M., Fowler J. W.. A multi-population genetic algorithm to solve multi-objective scheduling problems for parallel machines. *Computers & Operations Research*, 30(7) (2003), pp. 1087-1102.

[36] Kahn A. B.. Topological sorting of large networks. *Communications of the ACM*, 5(11) (1962), pp. 558-562.